\documentclass[aps,prb,twocolumn,superscriptaddress,nofootinbib,longbibliography]{revtex4-2}
\usepackage{graphicx,amsmath,amssymb,bm,dcolumn}
\usepackage{xcolor}
\newcommand{\new}[1]{#1}
\definecolor{revgreen}{rgb}{0,0.45,0.10}
\newcommand{\rev}[1]{#1}

\usepackage[colorlinks=true,linkcolor=blue,citecolor=blue,urlcolor=blue]{hyperref}

\begin{document}

\title{Certified reduced-basis emulation of conformal field theories on the fuzzy sphere}

\author{Virgil V. Baran}
\affiliation{Faculty of Physics, University of Bucharest,
077125 Bucharest-M\u{a}gurele, Romania}
\affiliation{``Horia Hulubei'' National Institute of Physics and
Nuclear Engineering, 077125 Bucharest-M\u{a}gurele, Romania}

\date{\today}

\begin{abstract}
Fuzzy-sphere regularization gives numerical access to three-dimensional conformal field theories (CFTs), but locating one requires testing for criticality across the parameter space of microscopic couplings. We reduce the search cost by about two orders of magnitude, combining an exact diagonalization based on states having definite total angular momentum with a certified reduced-basis emulator that reconstructs energies, observables and their derivatives continuously across the couplings from a handful of exact solutions per system size.  The combination reproduces the published ground-state-energy analysis of the fuzzy-sphere Ising model and the published conformal data of a three-flavor fuzzy-sphere realization of the O(2) Wilson--Fisher CFT, including quantities defined by parameter derivatives.  Applied to the O(2) model with no CFT input, the ground-state-energy criterion locates the critical coupling to within half a percent of a density matrix renormalization group value obtained on systems more than twice the size, and maps the critical line of the model across the coupling plane. Ground-state criteria locate the critical line robustly, but the point they select along it is weakly determined, and the computed dimensions come closest to the bootstrap elsewhere along the line.
\end{abstract}

\maketitle

\section{Introduction}

Conformal field theories (CFTs) govern the critical behavior at continuous
phase transitions, from the liquid--vapor and Ising transitions to the superfluid
and magnetic transitions of the O($N$) Wilson--Fisher family.  Nonperturbative access
to their operator content has traditionally come from lattice Monte
Carlo and, more recently, from the conformal bootstrap, which provides
rigorous determinations of low-lying scaling dimensions for the Ising
and O($N$) models \cite{Kos2016,Chester2020}.  The fuzzy-sphere regularization introduced by Ref.~\onlinecite{Zhu2023} has emerged as a complementary microscopic route in
which interacting fermions in the lowest Landau level (LLL) of a
monopole two-sphere realize a 3d CFT with radial quantization built in,
so that, by the state--operator correspondence, each eigenstate of the
critical Hamiltonian maps to a scaling operator with
$\Delta_\mathcal{O}\propto E_\mathcal{O}-E_0$.

The scope of the method has widened rapidly beyond the 3d Ising CFT
\cite{Zhu2023} to include the O($N$) Wilson--Fisher family through
bilayer models \cite{Han2024}, flavored fermions \cite{Guo2026}, quantum
rotors \cite{Dey2026b}, and spin-1 models \cite{Dey2026a}, as well as
free and interacting fermionic CFTs \cite{Zhou2026}, anyonic
realizations \cite{Voinea2025}, Chern--Simons-matter CFTs \cite{Zhou2026b},
and, via a fuzzy three-sphere,
four-dimensional criticality \cite{Meng2026}; see Ref.~\onlinecite{HeZhu2026}
for a recent review.

While small systems already display remarkably accurate conformal data, the coupling of irrelevant operators present in any microscopic realization leads to unavoidable finite-size effects. Their magnitude depends on the location in the parameter space
of the microscopic model, with different operators converging
at markedly different rates \cite{Dey2026a}.  The established criticality tests include the finite-size scaling of an order
parameter or of a Binder cumulant \cite{Zhu2023},  two-point
function matching \cite{Han2023}, conformal perturbation theory on the lowest
primary and its descendant \cite{Dey2026a}, or minimizing the distance
between the computed spectrum and the operator content of the target theory \cite{Guo2026,Meng2026}.

  Ref.~\onlinecite{Wiese2026} recently showed
that a remarkably economical test exists that needs nothing beyond the
ground-state (GS) energy.  At the critical point, the
size dependence of the GS energy acquires a universal nonanalytic piece, the Casimir energy of the CFT on $S^2\times\mathbb{R}$, which can be read off by fitting
\begin{equation}
\frac{E_{\rm GS}}{N}=E_0+E_1x+E_{3/2}\,x^{3/2}+E_2x^2+E_{5/2}x^{5/2},
\label{eq:fit}
\end{equation}
with $x=1/N$, across system sizes $N$.  The normalized coefficient
$\chi=E_{3/2}/E_0$ of the nonanalytic term is stationary exactly on the
critical manifold, so criticality can be located by mapping $\chi$ over the parameter space and finding where it is stationary transverse to the manifold. Along the manifold itself $\chi$ still varies, and its stationary point there is the operating point the criterion recommends, with the curvature around it encoding the leading corrections to scaling.  This requires evaluating $\chi$ on a dense grid
of couplings at every system size, which for the Ising study of
Ref.~\onlinecite{Wiese2026} meant on the order of a thousand exact diagonalization (ED) ground states
per size.  

Furthermore, the ED itself is commonly
performed in bases that fix only the angular-momentum projection $L_z$ (``$m$-scheme'' bases)
\cite{Wiese2026,Guo2026,Meng2026,Dey2026a,Zhou2025}. These are far larger than necessary, since states of every total angular
momentum $L$ are mixed together, and $L$ itself is assigned by
measuring $\langle L^2\rangle$ in each eigenstate, which
becomes ambiguous whenever levels of different $L$ happen to lie close
in energy.  This is the same
$m$-scheme versus $J$-scheme dichotomy long familiar in nuclear
configuration-interaction (shell-model) calculations
\cite{Caurier2005}, where $m$-scheme codes such as KSHELL
\cite{Shimizu2019}, BIGSTICK \cite{Johnson2018}, and ANTOINE
\cite{Caurier1999} dominate because Slater determinants make the
Hamiltonian sparse and simple, while angular-momentum-coupled
($J$-scheme) codes such as NuShellX \cite{Brown2014} and NATHAN \cite{Caurier1999b} attain far smaller
dimensions at the price of recoupling algebra. 

We address both bottlenecks here, and we first construct an exact-$L$ ED whose symmetry blocks are smaller by one to two orders of magnitude than the blocks diagonalized by the
$m$-scheme codes of Refs.~\onlinecite{Wiese2026,Guo2026,Dey2026a}
(see Fig.~\ref{fig:dims} below), with every state carrying exact quantum numbers. 

We then bring projection-based emulation, known as eigenvector continuation in nuclear physics \cite{Frame2018} and, more generally, as certified reduced-basis methods (RBM) \cite{RBbook,Duguet2024}, to fuzzy-sphere CFT scans.  Due to the extremal energy
eigenstates sweeping out a low-dimensional manifold as the couplings are
varied, the ground state at any point is well represented by a short
linear combination of exact solutions (snapshots) taken across the
parameter space.  The output is a continuous, interpolation-free energy surface (with parameter derivatives in closed form) that is built from just a few tens of
snapshots at fixed system size, rather than thousands of grid solves. 

Our physics results concern the O(2) (XY) CFT, realized on the fuzzy
sphere by a three-flavor (spin-1) model \cite{Dey2026a}.  Without any CFT input, the GS-energy criterion locates the critical coupling to within $0.5\%$ of the value Ref.~\onlinecite{Dey2026a} obtains at more than twice the system size, and maps the critical line of the model.  The $\chi$ extremum point the criterion selects along that line is far less sharply defined, and a direct spectral test finds the best agreement with the bootstrap at other couplings.

The remainder of this paper is organized as follows.
Section~\ref{sec:methods} defines the models, the exact-$L$ basis,
and the certified emulator.  Section~\ref{sec:ising} presents the Ising
validation and the conventions on which the $\chi$ analysis depends, and
Section~\ref{sec:o2} contains the O(2) results. The final Section~\ref{sec:concl} contains a summary and a discussion on possible extensions.
\section{Models and methods} \label{sec:methods} \subsection{Models} A
CFT in flat space can equivalently be formulated in radial quantization
on a sphere $S^2$ of radius $R$ times a time direction, where
dilatations play the role of time evolution and the dilatation operator
becomes the Hamiltonian.  Its eigenvalues
are then the scaling dimensions of the local
operators of the theory, $E_\mathcal{O}-E_0=(v/R)\,\Delta_\mathcal{O}$,
with $v$ a nonuniversal speed.  Any microscopic model that sits at a critical point in the right
universality class and is defined on a sphere with the rotational
symmetry \emph{exactly} preserved therefore has a spectrum that
organizes itself into conformal multiplets, each a primary of dimension
$\Delta$ accompanied by descendants at $\Delta+1,\Delta+2,\dots$ in
prescribed angular-momentum representations. 

The latter condition is incompatible with a lattice discretization of a sphere,
which breaks rotations to a finite subgroup; the resulting anisotropy is visible
in the spectrum at accessible sizes, where states within one angular-momentum
multiplet fail to remain degenerate \cite{Brower2013}, and recovering continuum
data requires tuned counterterms \cite{Brower2021} or an explicit accounting of
the rotation-breaking operators in the effective Hamiltonian
\cite{Lao2023,Wu2026}.  The fuzzy-sphere construction \cite{Zhu2023} evades this by discretizing
the sphere in a way that keeps SO(3) exact. Electrons on a sphere
threaded by a magnetic monopole of strength $2s$ have a lowest Landau
level of exactly $N=2s+1$ degenerate orbitals, which form a single
spin-$s$ irreducible representation.  Projecting the dynamics into that level
gives a finite Hilbert space on
which rotations act exactly, with no residual anisotropy at any
$N$.  The number of orbitals plays the role of area, $R\propto
\sqrt{N}$, and the continuum limit is $N\to\infty$.  In this context one gives up locality in the usual lattice sense, as the
built-in local interaction is smeared over a magnetic length by the LLL-projection, which is
why the construction is often described as a fuzzy, or noncommutative,
sphere.  In exchange, finite-size corrections are governed by the CFT's own irrelevant operators instead of by lattice artifacts, and are correspondingly small, so that systems of a few tens of orbitals already
reproduce bootstrap dimensions at the percent level.

Here we consider models of $N_f$-flavor fermions in the LLL of a monopole sphere with
$2s=N-1$, with $N$ orbitals accommodating $N$ particles interacting through an LLL-projected $H_{\rm int}$
built on the two-body kernel $U$ of Eq.~\eqref{eq:hint} below.
The $N_f=2$ Ising model's interaction couples only the two flavors' densities
\cite{Zhu2023},
\begin{equation}
H_{\rm int}^{\rm Ising}=2\sum_{\{m_i\}}U^{m_1m_2}_{m_4m_3}\,
c^\dagger_{\uparrow m_1}c^\dagger_{\downarrow m_2}
c^{\phantom{\dagger}}_{\downarrow m_3}c^{\phantom{\dagger}}_{\uparrow m_4},
\label{eq:hising}
\end{equation}
while the $N_f=3$ O(2) model's interaction is the flavor-scalar density--density
repulsion minus an XY exchange \cite{Dey2026a},
\begin{align}
H_{\rm int}^{\rm O(2)}=\sum_{\{m_i\}}U^{m_1m_2}_{m_4m_3}
\Bigl[&(c^\dagger_{m_1}\!\cdot c^{\phantom{\dagger}}_{m_4})
(c^\dagger_{m_2}\!\cdot c^{\phantom{\dagger}}_{m_3})
\nonumber\\[-2pt]
-\tfrac12\!\!\sum_{a=x,y}&(c^\dagger_{m_1}S_a c^{\phantom{\dagger}}_{m_4})
(c^\dagger_{m_2}S_a c^{\phantom{\dagger}}_{m_3})\Bigr],
\label{eq:ho2}
\end{align}
where $c_m$ collects the three flavors and $S_{x,y}$ are the spin-1
matrices; the second term is the ferromagnetic $xy$ exchange that drives the
ordering, and is the only piece that distinguishes flavor directions, since the density--density term is a flavor scalar. 

Both models share the LLL-projected two-body kernel
\begin{equation}
U^{m_1m_2}_{m_4m_3}=\sum_{\ell\ge0}V_\ell\sum_{M}
\langle s\,m_1;\,s\,m_2|J_\ell\,M\rangle
\langle s\,m_4;\,s\,m_3|J_\ell\,M\rangle,
\label{eq:hint}
\end{equation}
with $J_\ell=2s-\ell$; the two models differ only in the flavor
structure that dresses this kernel.  Here $V_\ell$ weights the channel
of relative angular momentum $\ell$: $V_0$ penalizes two particles at the same
point, $V_1$ penalizes the next-closest configuration, and so on; these
are the Haldane pseudopotentials of the quantum Hall problem. Truncating at $\ell\le1$ keeps the shortest-range repulsion, and by
Fierz-like rearrangement the same interaction contains both a
flavor-symmetric density--density piece and terms that exchange flavor
between the two particles.

The full $N_f=2$ Ising model also involves an additional transverse field,
\begin{equation}
H_{\rm Ising}=H_{\rm int}^{\rm Ising}(V_0,V_1{=}1)-h\sum_m
(c^\dagger_{\uparrow m}c^{\phantom{\dagger}}_{\downarrow m}+{\rm h.c.}),
\end{equation}
and realizes the Ising CFT along the critical line in $(V_0,h)$ seen in Fig.~\ref{fig:ising} below, whereas
the $N_f=3$ (spin-1) O(2) model \cite{Dey2026a} instead involves a single-ion anisotropy,
\begin{equation}
H_{\rm O(2)}=H_{\rm int}^{\rm O(2)}(V_0,V_1{=}1) +D\sum_m c^\dagger_m S_z^2 c^{\phantom{\dagger}}_m ,
\label{eq:ho2full}
\end{equation}
and realizes the O(2) CFT along a critical line in the $(V_0,D)$ plane.

In both cases the flavor-symmetric $H_{\rm int}$ sets the energy scale and supplies the interactions that make the transition continuous, while the remaining flavor-breaking one-body terms drive it.  In the Ising model the transverse field $h$ competes directly with the interaction.  At small $h$ the repulsion favors the pseudospins
aligning along $\sigma^z$, spontaneously breaking the $\mathbb{Z}_2$
that flips them, and the system is a quantum-Hall Ising ferromagnet; at
large $h$ every pseudospin polarizes along $\sigma^x$ and the ground
state is a unique, symmetric paramagnet.  Between the two lies a
continuous transition in the 3d Ising class.  In the O(2) model the
role of the field is played by the single-ion anisotropy $D$, which
raises the energy of the $S^z=\pm1$ flavors relative to $S^z=0$: for small $D$ the $xy$ exchange aligns the spins in the plane, giving
$\langle S^+\rangle\neq0$ and breaking the O(2) rotation of that plane,
while for large $D$ every orbital is driven into $S^z=0$, giving a
unique disordered state.    The transition between them is the O(2) Wilson--Fisher (XY)
fixed point.  The two models therefore share a structure in which an ordered phase
spontaneously breaks a global symmetry (the $\mathbb{Z}_2$ exchanging the two pseudospin flavors, and the O(2) rotation generated by the conserved charge $Q$) and a disordered phase leaves it intact.
Either symmetry survives the regularization as an exact quantum number
of the lattice Hamiltonian, and in each case the order parameter $\sigma$ is the lowest-dimension operator
charged under it.

In both cases $V_1=1$
fixes the overall energy scale, so $(V_0,h)$ and $(V_0,D)$ are complete
two-parameter families.

\subsection{Exact-$(L,Q)$ diagonalization}
\label{sec:jscheme}

\begin{figure*}[tb]
\includegraphics[width=\textwidth]{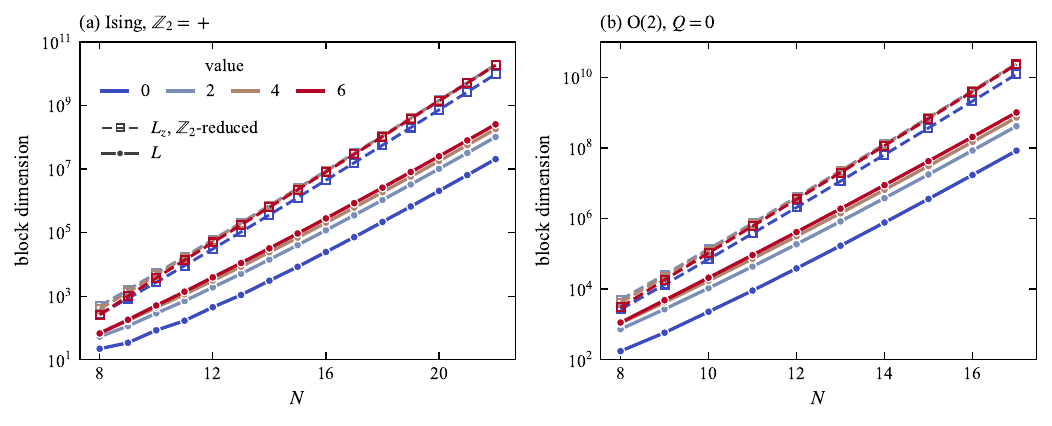}
\caption{\label{fig:dims}Symmetry-block dimensions versus system size for
blocks of fixed $L_z$ (dashed, open squares) and of fixed total $L$
(solid, filled circles), at the values $0,2,4,6$.  (a) Ising, at $\mathbb{Z}_2{=}+$.  (b) O(2), at $Q{=}0$.
Curves show the block each scheme actually diagonalizes.  Dashed: the
$(L_z,Q)$ block further reduced by the discrete $\mathbb{Z}_2$
symmetries implemented in FuzzifiED \cite{Zhou2025}, namely charge conjugation throughout $Q=0$ and the $\pi$-rotation $R_y$ within $L_z=0$.  Solid: the full exact-$(L,Q)$ block, which we do not reduce by
$C$ since the solver projects inside it.}
\end{figure*}

To solve the above models by exact diagonalization, we construct symmetry-adapted bases carrying \emph{exact} total angular
momentum $L$ and the flavor quantum number, rather than $L_z$ and the
Cartan charges alone.  We write these as exact-$(L,Q)$ blocks, $Q$ being
the O(2) charge; for the Ising model the flavor label is instead the
$\mathbb{Z}_2$ parity, and we abbreviate to exact-$L$ wherever it is the
rotational resolution that is at issue.
Fig.~\ref{fig:dims} compares block dimensions across sizes and
sectors for both models.  For O(2) the exact-$(L{=}0,Q{=}0)$ block is smaller than the $m$-scheme
block by a factor growing from $15$ at $N=8$ to $124$ at $N=16$, where
the $m$-scheme block, the one a FuzzifiED \cite{Zhou2025} run diagonalizes, is
$(L_z{=}0,Q{=}0)$ reduced by charge conjugation and by the
$\pi$-rotation $R_y$.  On $L_z=0$ states the latter has eigenvalue $(-1)^L$, so it recovers only the parity of $L$, whereas the exact-$(L,Q)$ construction resolves $L$ itself.  Since CFT spectroscopy needs only low $L$, what matters operationally is
that the largest block we ever diagonalize for a low-lying multiplet is
orders of magnitude below the $m$-scheme equivalent, and that every state carries
exact quantum numbers, with no a posteriori $\langle L^2\rangle$
assignment, which becomes ambiguous in near-degenerate multiplets.  For the Ising model the
blocks are small enough that the $N=17$ ceiling of
Ref.~\cite{Wiese2026} corresponds to only $7\times10^4$ states in our
scheme, and $N=20$--$21$ ($2.1$ and $6.5\times10^6$ states)
are directly accessible.

Our construction follows the $J$-coupled proton--neutron factorization of
NuShellX \cite{Brown2014} and NATHAN \cite{Caurier1999b}; it is
clearest in the Ising model, which maps onto a nuclear shell-model
problem exactly.  Rotating to
the $\sigma^x$ basis makes the transverse field diagonal, and the
two-flavor LLL problem becomes  a one-species,
$N$-particle system in two orbits of equal angular momentum $s=(N-1)/2$ and opposite
parity, carrying two single-particle energies $\pm h$ and four
two-body matrix elements at any size.  
The shell-model code's total angular momentum $J$ is the CFT spin
$L$, and its parity is the Ising $\mathbb{Z}_2$.  The basis is
then precisely NuShellX's,
\begin{equation}
|B,L\rangle=\bigl|\,[\,(n_+J_+\alpha_+)\otimes(n_-J_-\alpha_-)\,]\;L\,
\bigr\rangle ,
\label{eq:basis2}
\end{equation}
where $n_\pm$ is the number of particles in the orbit of parity $\pm$.
These are not separately conserved, since only $n_++n_-=N$ and the $\mathbb{Z}_2$ parity $(-1)^{n_-}$ are, so a block collects every
partition of the right parity, the $n_\pm$ particles of each orbit
coupling to good angular momentum $J_\pm$, with $\alpha_\pm$
distinguishing states that share the same $(n_\pm,J_\pm)$.  We build those states as highest-weight vectors, taking the kernel of
$L_+$ on the fixed-$L_z$ determinants that Gosper's algorithm enumerates
over occupation bitmasks.  We call each such angular-momentum-coupled group of one orbit's
particles a \emph{tower}.

In practice FuzzifiED is faster below $N\simeq13$, where our orbit-table
construction dominates the cost.  The crossover falls near $N=14$, and
by $N=16$ the exact-$L$ solver leads by $7$--$20\times$: $10.9$\,s
against $78$\,s in $(L,\mathbb{Z}_2)=(0,+)$, where its block is
$180\times$ smaller than the $m$-scheme one, and $28$\,s against
$567$\,s in $(2,+)$, where it is $74\times$ smaller.

\rev{The three-flavor O(2) model differs in two
ways.}  First, the basis acquires a third tower, coupled as
\begin{equation}
|B,L\rangle=\bigl|\,[\,(J_+\,J_-)\,J_c\,,\,J_0\,]\;L\,\bigr\rangle ,
\quad \sum_{\tau=\pm,0} n_\tau=N ,
\label{eq:basis}
\end{equation}
with the flavor charge $Q=n_+-n_-$ fixed blockwise, so that $L$ and $Q$
are both exact by construction.  Each interflavor term is brought to particle--hole (Pandya) form by
recoupling.  Write $A^\dagger_{J_0M}=[c^\dagger_\tau c^\dagger_{\tau'}]^{J_0}_M$, $\tau\neq\tau'$, for the pair creation operator and
$T^{\lambda}_{\mu}(\tau)=[c^\dagger_\tau\tilde c_\tau]^{\lambda}_{\mu}$ for the
one-body density tensor of flavor $\tau$, where $\tilde c_{\tau m}=(-1)^{s-m}c_{\tau,-m}$ is the time-reversed annihilation operator,
which transforms as a spherical tensor of rank $s$.  All four
single-particle labels equal $s$, and $J_0=2s$.  Then
\begin{align}
\sum_M A^\dagger_{J_0M}A_{J_0M}&=\sum_\lambda c_\lambda
\sum_\mu(-1)^{\mu}\,T^{\lambda}_{\mu}(\tau)\, T^{\lambda}_{-\mu}(\tau') ,
\nonumber\\[-2pt]
c_\lambda&=(2J_0+1)
\begin{Bmatrix} s & s & J_0\\ s & s & \lambda\end{Bmatrix} .
\label{eq:pandya}
\end{align}
The coefficient in Eq.~\eqref{eq:pandya} is the standard Pandya
coefficient \cite{Brown2014} specialized to a single $s$ shell and a single pair channel, which is why it reduces to one $6j$ symbol with no sum, and, as there, applies only between distinct flavors.

With the interaction strength
included, $F_\lambda=2V_0\,c_\lambda$, a matrix element between basis
states \eqref{eq:basis} factorizes into a product of \emph{per-flavor}
reduced density matrices,
\begin{equation}
\rho_\tau(\lambda)=\bigl\langle n_\tau J_\tau^{f}\alpha^{f}\bigr\|
[c^\dagger_\tau\tilde c_\tau]^{\lambda} \bigl\|n_\tau
J_\tau^{i}\alpha^{i}\bigr\rangle ,
\label{eq:rdm}
\end{equation}
tied by a single recoupling coefficient,
\begin{equation}
\langle B_f,L|H_{\tau\tau'}|B_i,L\rangle= \sum_\lambda
\Gamma_\lambda\,F_\lambda\; \rho_\tau(\lambda)\,\rho_{\tau'}(\lambda),
\label{eq:fact}
\end{equation}
where $\Gamma_\lambda$ is the $9j$ symbol coupling the initial and final
tower momenta to $L$, which reduces to a $6j$ because one of its
arguments vanishes.  The reduced density matrices \eqref{eq:rdm} are computed once per flavor tower (objects of small dimension) and reused for every one of the far more numerous coupled states; this reuse
is the source of the $J$-scheme's economy.  Matrix elements
are formed on the fly during each Lanczos multiplication, so only the
tower-level tables and the coupled vectors are stored.

The second difference is that the $V_1$ channel transfers a pair between
flavors and so does not conserve the $n_\tau$ separately.  Alongside the density tables \eqref{eq:rdm} we
therefore tabulate pair-creation reduced matrix elements
$\langle n_\tau{+}2,J'\|[c^\dagger_\tau c^\dagger_\tau]^{J_0}\|
n_\tau,J\rangle$, and Eq.~\eqref{eq:fact} acquires a companion in
which the two flavor factors change particle number oppositely.  The
three-tower coupling also requires one further recoupling relative to
Eq.~\eqref{eq:basis2}, since $H_{+-}$ acts inside $(J_+J_-)J_c$ while
$H_{\pm0}$ couples across the intermediate.

\emph{Charge conjugation.}  The O(2) group is $U(1)\rtimes\mathbb{Z}_2$,
and the $\mathbb{Z}_2$ of charge conjugation, $C:Q\to-Q$, is not diagonalized by $(L,Q)$ alone.  At $Q\neq0$ it maps a block to its partner
at $-Q$ and so carries no extra label, but at $Q=0$ it acts \emph{within}
the block and splits it.  On the basis \eqref{eq:basis} it exchanges the
two charged towers,
\begin{align}
C\,\bigl|[(J_+J_-)J_c,J_0]\,L\bigr\rangle
&=(-1)^{n_+n_-}(-1)^{J_++J_--J_c}
\nonumber\\[-2pt]
&\quad\times\bigl|[(J_-J_+)J_c,J_0]\,L\bigr\rangle ,
\label{eq:cop}
\end{align}
the first sign from reordering the two blocks of fermions and the second
from the recoupling $[(ab)c]\to[(ba)c]$, so $C$ is a signed permutation
and a symmetric involution. 

Of the operators needed at $Q=0$ the identity and $\varepsilon$ are
$C$-even, $j$ is $C$-odd and $T$ is $C$-even, so $(0,0)$ and $(1,0)$
deliver the wanted state as the lowest of their block, while $(2,0)$
does not, because there the $C$-odd descendant $\partial j$ and $T$
cross along the critical line at finite $N$ (opposite $C$ forbids level
repulsion).

Rather than build $C$-adapted blocks we project the Lanczos start vector
and every matrix--vector product onto the target parity, with the
complementary subspace shifted above the spectrum so it cannot be
mistaken for a ground state.  The stress tensor is then the lowest state
of its own sector, and a single Lanczos pair suffices.  Symmetrizing the
basis instead would halve the block, but $C$ pairs each channel with its
$J_+\leftrightarrow J_-$ partner and only the $J_+{=}J_-$ channels are self-paired ($16\%$ of the $(2,0)$ block at $N=10$, $11\%$ at $N=12$), so the symmetric combinations run across channel pairs,  every entry of the task table would acquire a $C$ partner, and the factorization \eqref{eq:fact} itself would have to be rewritten.  Projection instead keeps the block at full dimension and so costs roughly a factor of two.

\emph{Verification of the algebra.}  Every channel weight is projected
numerically from the corresponding $m$-scheme coefficient tensor with an
asserted residual, and every phase convention is fitted against
explicitly constructed Clebsch--Gordan states rather than taken from a
formula.  All $6j$ symbols are evaluated in exact rational arithmetic
before conversion to floating point; the alternating Racah sum otherwise
cancels catastrophically at the angular momenta reached here.  The solver is validated level by level against
direct $m$-scheme diagonalization at $N=4,6,8$ to $10^{-13}$ in every
sector, and against FuzzifiED \cite{Zhou2025} wherever both run; the
sparse matrix--vector product is checked separately against a reference
implementation on random vectors.

\subsection{Certified greedy reduced basis}
\label{sec:rbm}

For problems such as scanning a parameter space with ED, projection-based emulators, known as eigenvector continuation (EC) in quantum many-body physics
\cite{Frame2018} and as (certified) reduced-basis methods (RBM) in the
numerical-analysis literature \cite{RBbook}, replace the full solve
by a small projected one; see Ref.~\onlinecite{Duguet2024} for a recent
review.  These methods are by now standard tools in nuclear-physics
uncertainty quantification and they have been applied to equilibrium condensed matter systems \cite{Herbst2022Apr, Baran2023Apr, Brehmer2023Aug} and recently generalised to open quantum systems \cite{Christiansen2025}, but have not, to our knowledge, been applied
to fuzzy-sphere CFT scans. 

The fuzzy sphere natively meets the structural requirements of the reduced-basis construction, which makes certified emulation particularly effective here. First, fuzzy-sphere Hamiltonians are exactly \emph{affine} in their couplings, since Haldane pseudopotentials and one-body fields enter linearly,
\begin{equation}
H(\bm\theta)=\mathcal{H}_0+\sum_{i=1}^{p}\theta_i\,\mathcal{H}_i ,
\label{eq:affine}
\end{equation}
with parameter-independent operators $\mathcal{H}_q$ (here $p=2$, with
$\bm\theta=(V_0,h)$ or $(V_0,D)$).  In the general reduced-basis setting
an approximate affine form must be manufactured by empirical
interpolation \cite{RBbook}; here Eq.~\eqref{eq:affine} is exact, so
every projected quantity below (the emulated energy, its error certificate, and its derivatives with respect to $\bm\theta$) is exact as well.  Second, the exact-$(L,Q)$ basis of the previous subsection removes all
level crossings with other symmetry sectors from the scan; within one
sector the ground state evolves analytically in $\bm\theta$, \rev{and the RBM
then converges rapidly with the number of}
snapshots \cite{Duguet2024}.  Fig.~\ref{fig:isingconv} shows that
convergence directly, against exact solves, for one of the emulators used
below.

The emulator is built in an \emph{offline} stage.  Snapshots
\new{$|\varphi_a\rangle$}, the exact ground states at parameter points
$\bm\theta_a$ ($a=1\ldots r$), are computed by Lanczos in the
exact-$(L,Q)$ block and orthonormalized into the columns
$\psi_1,\ldots,\psi_r$ of a matrix $\Psi$.  The operators are projected once,
$\mathsf{h}_q=\Psi^\dagger\mathcal{H}_q\Psi$, along with the Gram tensor
$\mathsf{B}^{qq'}=(\mathcal{H}_q\Psi)^\dagger(\mathcal{H}_{q'}\Psi)$.
In the \emph{online} stage, for arbitrary $\bm\theta$ the $r\times r$
problem
\begin{equation}
\Bigl[\mathsf{h}_0+\sum_{i=1}^{p}\theta_i\,\mathsf{h}_i\Bigr]\,y=E\,y
\label{eq:reduced}
\end{equation}
is solved in $\mathcal{O}(r^3)$ time (microseconds at the ranks that occur in practice).  Because Eq.~\eqref{eq:reduced} is affine in $\bm\theta$, its
characteristic polynomial has degree $r$ with coefficients polynomial in
$\bm\theta$, so the emulated energies, and the dimension ratios built from
them, are algebraic functions of the couplings rather than interpolants of
sampled values.  \new{Its solution, normalized to $y^\dagger y=1$, defines the emulated ground state}
\begin{equation}
\new{|\Psi(\bm\theta)\rangle\equiv\Psi\,y(\bm\theta)
=\sum_{a=1}^{r}y_a(\bm\theta)\,\psi_a}
\label{eq:emustate}
\end{equation}
as a vector in the full sector space, where the $y_a$ are coordinates in the orthonormal basis.  Because
$\Psi^\dagger\Psi=\mathsf{1}_r$, the $r\times r$ identity,
$|\Psi(\bm\theta)\rangle$ is itself normalized and the lowest Ritz value is a variational
upper bound on the true ground-state energy.  The emulator also carries
its own error certificate, the exact residual norm of the Ritz pair
\new{$(E,|\Psi(\bm\theta)\rangle)$}
being available at $\mathcal{O}(r^2)$ cost,
\begin{equation}
\rho(\bm\theta)^2\equiv\bigl\|[H(\bm\theta)-E]\,\new{|\Psi(\bm\theta)\rangle}\bigr\|^2
= y^\dagger \mathsf{B}(\bm\theta)\,y-E^2 ,
\label{eq:resid}
\end{equation}
with $\mathsf{B}(\bm\theta)=\sum_{q,q'=0}^{p}\theta_q\theta_{q'}
\mathsf{B}^{qq'}$ and $\theta_0\equiv1$.  For Hermitian $H$ the distance
from $E$ to the spectrum is bounded by $\rho$, and the error of the
targeted level by $\rho^2/\delta$ with $\delta$ the gap to the rest of
the spectrum  \cite{Parlett}.  Snapshots are selected \emph{greedily}, with the next $\bm\theta_{r+1}$
chosen as the point of worst certified residual over a finite candidate
set, so the basis adapts itself to the solution manifold; the rank
saturates when the manifold is exhausted.  

\rev{In practice a spectral shift
$H\to H-\zeta$, with $\zeta$ the ground-state energy at the centre of the
scanned box, removes the $E^2$ cancellation floor of
Eq.~\eqref{eq:resid} when $|E|$ is large.}  Snapshot orthogonalization requires reorthogonalization (``twice is
enough'') together with an explicit $\|\Psi^\dagger\Psi-\mathsf{1}\|$
gate; a single Gram--Schmidt pass can lose orthogonality badly enough to
place the reduced ground state below the exact one, violating the
variational bound.   Every emulator is also
gated by off-snapshot exact solves.  \rev{Each new exact solve is warm-started from the reduced-basis}
prediction, which accelerates the Lanczos solves by $1.6$--$1.7\times$.  The construction generalizes directly to the $k$
lowest states per sector (multi-state EC \cite{Duguet2024}), with the
$k$th Ritz value a variational bound on the $k$th level by the
Poincar\'e separation theorem.

Because
Eq.~\eqref{eq:affine} is exact, Eq.~\eqref{eq:reduced} is an ordinary
$r$-dimensional Hermitian eigenproblem whose matrix depends linearly on
$\bm\theta$, and the derivatives of the Ritz value are available in
closed form from the eigendecomposition already performed.  The
Hellmann--Feynman theorem applied inside the reduced space gives the
gradient,
\begin{equation}
\frac{\partial E}{\partial\theta_i}=y^\dagger\mathsf{h}_i\,y ,
\label{eq:hf}
\end{equation}
and second-order perturbation theory in the same space, with $(E_m,y_m)$
the remaining Ritz pairs of Eq.~\eqref{eq:reduced}, gives the Hessian,
\begin{equation}
\frac{\partial^2 E}{\partial\theta_i\,\partial\theta_j}
=2\,\mathrm{Re}\sum_{n\neq0}
\frac{(y^\dagger\mathsf{h}_i\,y_n)(y_n^\dagger\mathsf{h}_j\,y)}{E-E_n} ,
\label{eq:hess}
\end{equation}
with no $\partial^2H/\partial\bm\theta^2$ term, $H(\bm\theta)$ being
linear. These are exact derivatives of the emulator and are essentially free, since the full set
$\{(E_m,y_m)\}$ comes from the single dense eigensolve of
Eq.~\eqref{eq:reduced}, so gradient and Hessian add only
$\mathcal{O}(p^2r^2)$ to the $\mathcal{O}(r^3)$ already spent.  Every
denominator in Eq.~\eqref{eq:hess} is negative for the lowest level, so
$E(\bm\theta)$ is concave, as a pointwise minimum of a family of
functions linear in $\bm\theta$ must be.  The construction needs only
that the reduced ground state remain nondegenerate, the same condition under which it evolves analytically within the sector.

At fixed window the fit of Eq.~\eqref{eq:fit} is a linear
map $\mathsf{c}=\mathsf{P}\,\mathsf{f}$ of the data
$\mathsf{f}_N=E_{\rm GS}(N,\bm\theta)/N$ onto the coefficient vector
$\mathsf{c}=(E_0,E_1,E_{3/2},E_2,E_{5/2})^{\mathsf{T}}$, with
$\mathsf{P}$ the pseudoinverse of the constant matrix
$\mathsf{X}_{Ns}=N^{-s}$ assembled from the exponents
$s=0,1,3/2,2,5/2$.  Since $\mathsf{P}$ carries no $\bm\theta$
dependence, the coefficient derivatives are that same map applied to the
derivatives of the data,
$\partial_i\mathsf{c}=\mathsf{P}\,\partial_i\mathsf{f}$, and the
quotient rule applied to $\chi=E_{3/2}/E_0$ delivers $\chi$,
$\nabla\chi$ and its Hessian in closed form at any $\bm\theta$, at the
cost of one reduced eigensolve per size in the window.  The stationary point
is then found by Newton's method on $\nabla\chi=0$, without a grid or a smoothing convention.  On the Ising surface this recovers the stationary
points obtained on the microgrid with the smoothing of
Ref.~\onlinecite{Wiese2026} to $4\times10^{-5}$ in $h^*$,
$9\times10^{-4}$ in $V_0^*$, and $0.05\%$ in $\chi^*$ and in the stiff
curvature.  Because one
gradient evaluation costs a handful of $r\times r$ eigensolves,
$|\nabla\chi|$ can further be mapped over an entire search box and each
local stationary point polished and classified.

Observables, such as order parameters needed here, follow the same offline--online split.  On a finite sphere
the order parameter itself averages to zero by symmetry, so the quantity that locates a transition is its second
moment, $\langle m^2\rangle=\|\mathcal{O}\,|{\rm GS}\rangle\|^2$ with
$\mathcal{O}$ the symmetry-odd operator, $\sum_m\sigma^z_m$ for Ising and its $Q$-raising analogue for O(2). We assemble, once,
\begin{equation}
\mathsf{G}=(\mathcal{O}\Psi)^\dagger(\mathcal{O}\Psi)
\label{eq:obs}
\end{equation}
from $r$ applications of $\mathcal{O}$ to the stored basis, after which
$\langle m^2\rangle(\bm\theta)=y^\dagger\mathsf{G}\,y$ costs nothing
beyond the reduced eigensolve already performed, at every point of the
box.

\begin{figure*}[ht!]
\includegraphics[width=\textwidth]{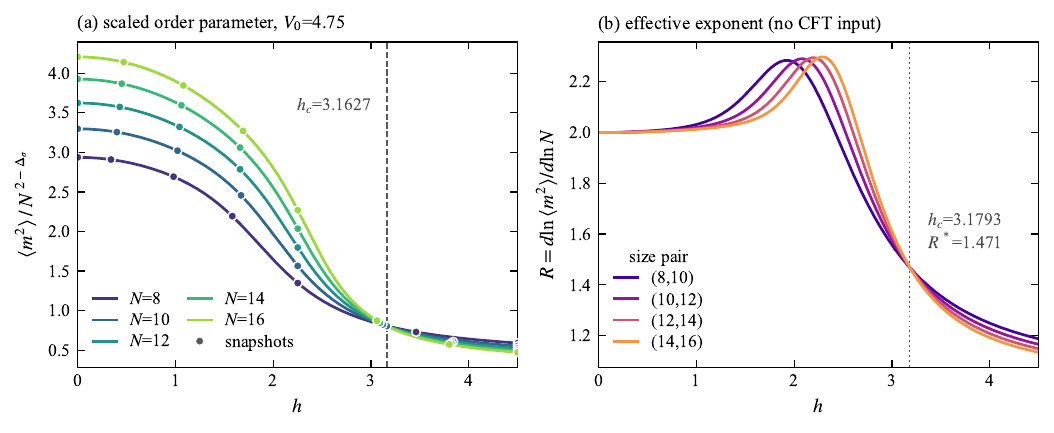}
\caption{\label{fig:isingop}Finite-size scaling of the Ising order
parameter at $V_0=4.75$, from certified emulators at $N=8$--$16$.  Here
$\langle m^2\rangle=\|\mathcal{O}\,|{\rm GS}\rangle\|^2$ with
$\mathcal{O}=\sum_m\sigma^z_m$ the $\mathbb{Z}_2$-odd order parameter,
which maps the $\mathbb{Z}_2{=}+$ block onto $\mathbb{Z}_2{=}-$.
Symbols mark the greedy snapshots (Sec.~\ref{sec:rbm})\new{, each emulator grown
until the relative change in $\langle m^2\rangle$ between consecutive snapshots
falls below $10^{-5}$: seven snapshots at $N=8$ and $10$, eight at $N=12$, $14$
and $16$, for sector dimensions $22$ to $24\,552$.}  (a) The scaled order parameter $\langle
m^2\rangle/N^{\,2-\Delta_\sigma}$, the estimator of
Ref.~\cite{Zhu2023}, which takes $\Delta_\sigma=0.5181489$
\cite{Kos2016} as input.  Consecutive sizes cross at $3.1512$, $3.1574$,
$3.1607$ and \new{$3.1627$}, the last from $(14,16)$ (dashed line).
At $h=0$ the ground state is fully $\mathbb{Z}_2$ ordered and $\langle
m^2\rangle=N^2$ exactly, which the construction returns at every size.  (b) The effective exponent $R=d\ln\langle m^2\rangle/d\ln N$
between consecutive sizes; the pair curves cross  $h_c=\new{3.1927}$,
\new{$3.1844$} and \new{$3.1793$} (dotted line), with crossing values
$R^*=\new{1.4637}$, $1.4678$, $1.4708$, drifting toward $2-\Delta_\sigma=1.4819$.}
\end{figure*} 

\begin{figure*}[tb]
\includegraphics[width=\textwidth]{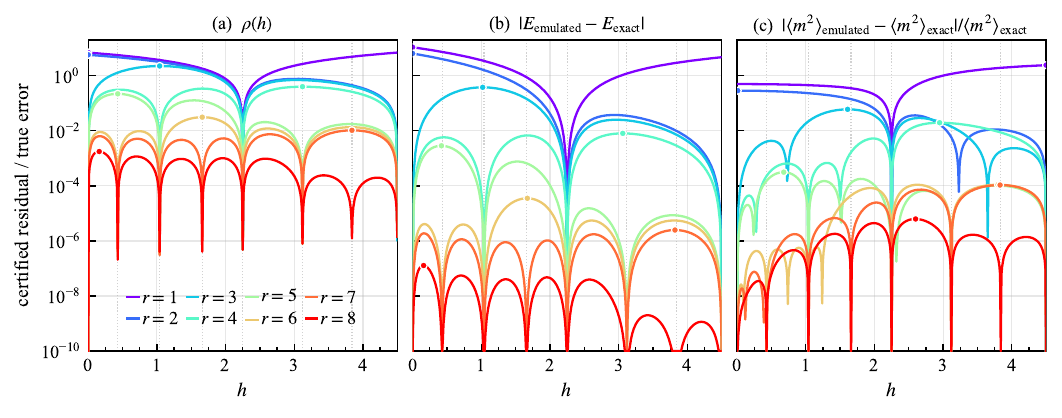}
\caption{\label{fig:isingconv}\rev{The certificate residual against actual
energy and order-parameter errors, for the $N=12$ Ising emulator on the cut of
Fig.~\ref{fig:isingop}, where the $(L,\mathbb{Z}_2)=(0,+)$ block has dimension
$448$; one colour per greedy iteration $r$.  (a) The certified residual
$\rho(h)$ of Eq.~\eqref{eq:resid}.  (b) The energy error
$|E_{\rm emulated}-E_{\rm exact}|$ and (c) the relative order-parameter error
$|\langle m^2\rangle_{\rm emulated}-\langle m^2\rangle_{\rm exact}|/\langle
m^2\rangle_{\rm exact}$, both against exact solves at every point of the same
grid.}}
\end{figure*}

\begin{figure}[ht!]
\includegraphics[width=\columnwidth]{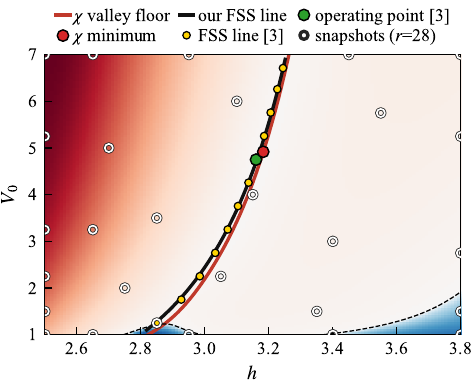}
\caption{\label{fig:ising}Heat map of
$\mathrm{sign}(\chi)\sqrt{|\chi|}$ over the $(h,V_0)$ plane at
$N^{\max}=16$ (sliding window $N=12$--$16$) from exact-$L$ ED +
greedy emulators.  Rings mark the $r=28$ snapshot locations that generate
the entire $N=16$ energy surface, selected over a $25\times27$ candidate grid
($\delta V_0=0.25$, $\delta h=0.05$).  The yellow circles indicate the
finite-size-scaling transition line digitized from Fig.~2(b) of
Ref.~\cite{Zhu2023} and the green
circle corresponds to their conventional operating point. Dashed lines separate the red and blue regions with positive and negative $\chi$. See also Fig. 4 of Ref.~\onlinecite{Wiese2026}. }
\end{figure}

\begin{figure*}[tb]
\includegraphics[width=\textwidth]{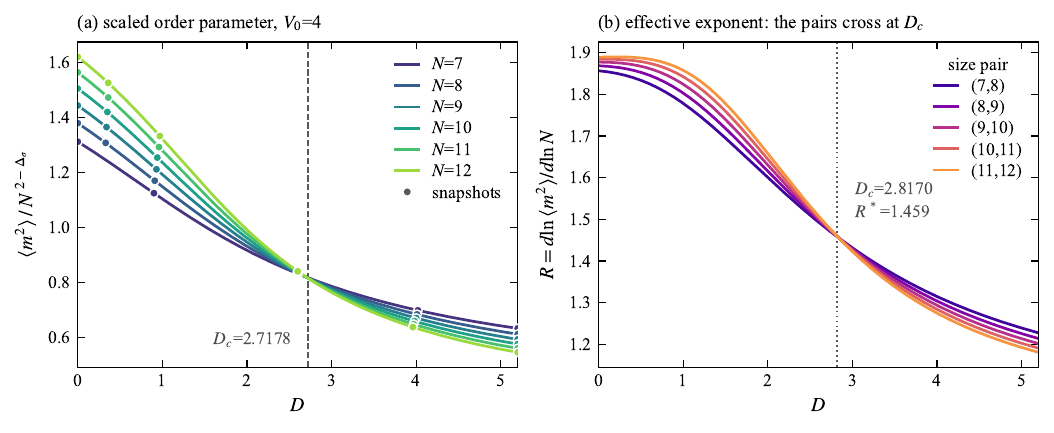}
\caption{\label{fig:o2op}Same as Fig.~\ref{fig:isingop} for the O(2) model, at
$V_0=4$ and $N=7$--$12$, with $\langle m^2\rangle$
defined in Sec.~\ref{sec:o2op}.  Symbols mark the greedy snapshots
of the cut emulators (Sec.~\ref{sec:rbm})\new{, each grown until the relative
change in $\langle m^2\rangle$ between consecutive snapshots falls below
$10^{-5}$: five snapshots at $N=7$ and six at every larger size, for sector
dimensions $50$ to $38\,587$.}  (a) The scaled order parameter $\langle
m^2\rangle/N^{\,2-\Delta_\sigma}$, with $\Delta_\sigma=0.519088$
\cite{Chester2020} as input: consecutive sizes cross at $D_c=2.6737$,
$2.6900$, $2.7019$, $2.7108$ and $2.7178$, the last from $(11,12)$
(dashed line).  (b) The effective exponent $R=d\ln\langle
m^2\rangle/d\ln N$, which uses no CFT input at all: consecutive pairs
cross at \new{$2.8575$}, \new{$2.8398$}, \new{$2.8269$} and \new{$2.8170$} (dotted line), with
crossing values $R^*=1.4522$, $1.4552$, $1.4573$, $1.4593$ drifting
toward $2-\Delta_\sigma=1.4809$.}
\end{figure*}

\section{Ising model validation}
\label{sec:ising}

For reproducing the finite size scaling results presented in Fig. 3 of Ref.~\onlinecite{Zhu2023} we build a family of one-dimensional emulators, one for each system size, that scan an $h$-interval at fixed $V_0$. In this Ising case, the $\mathbb{Z}_2$-odd operator $\mathcal{O}=\sum_m\sigma^z_m$ maps the
$\mathbb{Z}_2{=}+$ block onto $\mathbb{Z}_2{=}-$.

\begin{figure}[tb]
\includegraphics[width=\columnwidth]{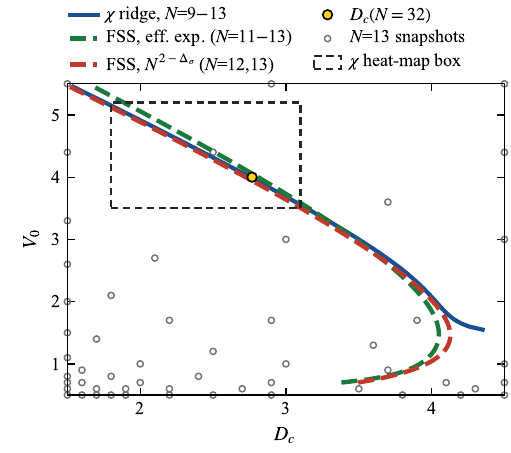}
\caption{\label{fig:o2line}The O(2) critical line from three finite-size estimators, all from
one set of emulators over the greedy box
$V_0\in[0.5,5.5]\times D\in[1.5,4.5]$, snapshots selected over a
$51\times31$ candidate grid ($\delta V_0=\delta D=0.1$).  Each size is grown
until a new snapshot is numerically linearly dependent on those already held; the certificate saturates near $3\times10^{-5}$.  Blue: the ridge of
$\chi=E_{3/2}/E_0$ on the window $N=9$--$13$, where the five coefficients of
Eq.~\eqref{eq:fit} are exactly determined, with no CFT input.  Green: the
order-parameter FSS line from the effective exponent, size pairs
$(11,12)\times(12,13)$, likewise with no CFT input.  Red: the same order
parameter through $N^{\,2-\Delta_\sigma}$ scaling at $(12,13)$. Circle: the DMRG
value $D_c(N{=}32)=2.7677$ at $V_0=4$ of Ref.~\onlinecite{Dey2026a}, their
largest-size result.  Open circles: the $54$ snapshots of the
$N=13$ emulator, the largest size entering any curve.  Dashed rectangle:
the window of Fig.~\ref{fig:o2map}.}
\end{figure}

\begin{figure}[!tp]
\includegraphics[width=\columnwidth]{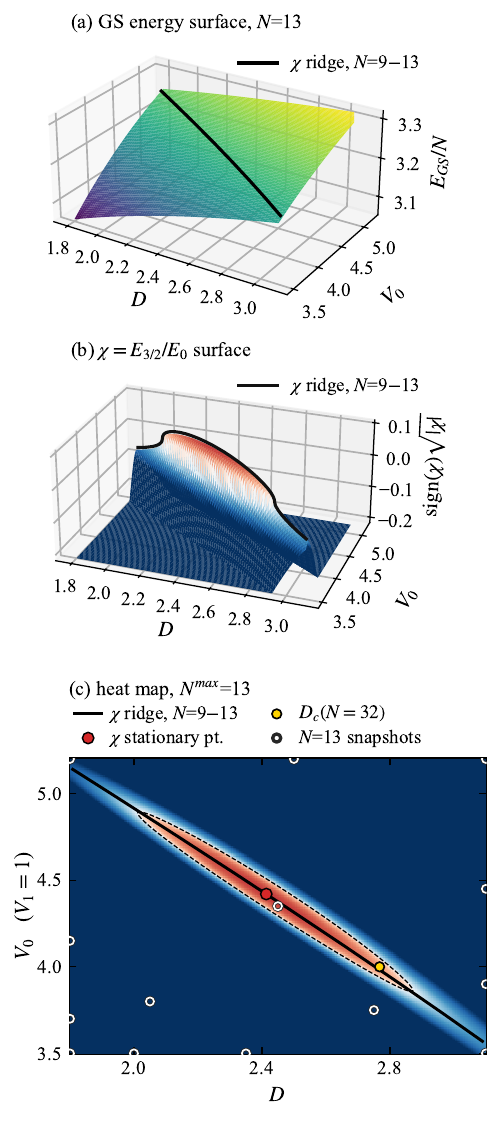}
\caption{\label{fig:o2map}The O(2) model over the $(D,V_0)$ plane at
$N^{\max}=13$.  (a) The GS energy surface is smooth and featureless.  The transition is invisible in $E$ itself.  (b) The same data reduced to
$\chi=E_{3/2}/E_0$, manifesting a sharp ridge.  (c) Heat map of sign$\chi \sqrt{|\chi|}$ with the
$D_c(N{=}32)$ of Ref.~\onlinecite{Dey2026a} at $V_0=4$ (yellow circle), greedily selected over a \new{$35\times27$
candidate grid ($\delta V_0=\delta D=0.05$).  Fourteen exact
solves generate the entire surface at this size}.  The window is the rectangle drawn on
Fig.~\ref{fig:o2line}.}
\end{figure}

Our Fig.~\ref{fig:isingop} shows the finite-size scaling that follows, at fixed 
$V_0=4.75$, obtained at each size by a one-dimensional emulator over the full field range of Ref.~\onlinecite{Zhu2023}. Fig.~\ref{fig:isingop}(a) uses their
estimator $\langle
m^2\rangle/N^{\,2-\Delta_\sigma}$, while panel (b) employs $R=d\ln\langle m^2\rangle/d\ln N$ which requires no CFT input. The two bracket $h_c$ and
close on it as the sizes grow, to \new{$3.1627$} and \new{$3.1793$} at the largest pair.  Note that the excursion of $R$  above
$2$ near $h\simeq2$ is a finite-size effect and not a
violation of $\langle m^2\rangle\le N^2$: it is $\langle
m^2\rangle/N^2$ approaching that bound from below faster than $N^2$
grows.  And at the upper end of the range $\langle m^2\rangle/N$ has
only reached $1.61$--$1.80$, so the disordered limit, where $R\to1$,
lies well beyond $h=4.5$.

We also show in Fig.~\ref{fig:isingconv} the decrease of the residual (\ref{eq:resid}), together with the corresponding energy and order parameter errors across the reduced basis construction of the one-dimensional $N=12$ emulator. The certificate is seen to comfortably bound the energy error at every $r$ and every $h$.

Carried across the $(V_0,h)$ plane the same crossings give a
critical \emph{line}, drawn in Fig.~\ref{fig:ising}. A single two-parameter emulator per size supplies
$\langle m^2\rangle$ everywhere, and the line is the locus of crossings
read off it.  It agrees with the line digitized from Fig.~2(b) of
Ref.~\onlinecite{Zhu2023}, within the accuracy of the digitization.

The emulator's continuous ground state energy surface allows us to apply the ground-state-energy criterion and reproduce the
analysis of Ref.~\onlinecite{Wiese2026} from our own data (exact-$L$ ED +
greedy emulators over the critical region of the $(V_0,h)$ plane,
$N\le18$).  Reproducing the published numbers
requires fixing a few conventions on which the results depend
sensitively:

\emph{(a) The fit window.} Eq.~\eqref{eq:fit} with the window
$N=8\ldots N^{\max}$ (least squares for $N^{\max}>12$) does not
reproduce the published minima, agreeing at $N^{\max}=12$ and departing
progressively beyond.  Agreement precisely where the
five-coefficient fit is exactly determined identifies the convention as
the sliding window $N\in[N^{\max}-4,\,N^{\max}]$, which is used
throughout below.

\emph{(b) The truncation order.}  Ref.~\onlinecite{Wiese2026} settles
on the five terms of Eq.~\eqref{eq:fit}.  Adding or dropping one describes
$E_{\rm GS}/N$ equally well, so the data do not themselves select an
order; $\chi$ nevertheless changes by more than $100\%$, and the
stationary point shifts by several tenths in $V_0$ or disappears
altogether.  The critical line barely moves under the same change, so the two parts of the criterion are not equally reliable, as the truncation leaves the line itself almost untouched, while the point along it where $\chi$ is stationary is not fixed by the data at any order reachable with $N\le18$.

Overall the $\chi$ valley agrees well with the order-parameter lines of Fig.~\ref{fig:ising}. The valley drifts monotonically toward the FSS
line as the window grows, and a power-law extrapolation in $N^{\max}$
closes it.  Its origin is the $o=4$ truncation of Eq.~\eqref{eq:fit},
which absorbs the neglected higher corrections into $E_{3/2}$ with
window-dependent weight and so displaces the valley to larger $h$, the
same displacement reported in Ref.~\onlinecite{Wiese2026}.

\section{O(2) model}
\label{sec:o2}

\label{sec:o2op}
\rev{In the $O(2)$ model the order parameter
carries} $Q=1$ and maps the $(0,0)$ block onto $(0,1)$, and since $\langle
m\rangle$ vanishes on a finite sphere by symmetry the measurable quantity
is $\langle m^2\rangle=\|\mathcal{O}\,|{\rm GS}\rangle\|^2$.
Fig.~\ref{fig:o2op} shows the two estimators at $V_0=4$, with consecutive
sizes crossing at $D_c=2.7178$ through the scaled order parameter and at
\new{$2.8170$} through the effective exponent, bracketing the $2.7677$ that
Ref.~\onlinecite{Dey2026a} reaches by DMRG at $N=32$.

The ground-state-energy criterion, applied along the same cut with
one-parameter emulators over $D$ at $N=8\ldots14$, gives a
$\chi$-stationary point that rises from $2.7522$ to $2.7549$ as the window
grows, with no CFT data used at any stage, and at $0.46\%$ from the DMRG
value. Ref.~\onlinecite{Dey2026a}
needs sizes out to $N=32$ to reach it, its $N=14$ estimate still standing at
$2.7985$.
The same emulators also reproduce that reference's own critical couplings and OPE
coefficients at every shared size, as detailed in 
Appendix~\ref{app:ope}.

Carried across the $(V_0,D)$ coupling plane the same three constructions each give a
critical \emph{line}, drawn in Fig.~\ref{fig:o2line} over
$V_0\in[0.7,5.5]$.  They are independent, two reading the order parameter
and one the ground-state energy alone, and they agree closely above
$V_0\simeq2$. We are not aware of this line having been mapped before, with Ref.~\onlinecite{Dey2026a} working throughout at the single coupling $V_0=4$, and the O($N$) phase diagram of Ref.~\onlinecite{Guo2026} being schematic.  

The O(2) line displays two features  with no Ising counterpart.  First, in
Fig.~\ref{fig:o2line} all three estimators rise steeply toward small $V_0$,
and the two order-parameter lines then turn over inside the box, around
$D=4.1$, $V_0=1.5$, the only place where the line is not monotone.  One would expect
$D_c$ to rise with $V_0$, since $V_0$ is a repulsion and raising it lifts
the overall energy scale against which $D$ must compete.  Over most of the
range it falls instead, because the $\ell=0$ channel does not act on the
two phases alike.  That channel carries the maximal pair angular momentum
$J_0=2s$ and is therefore spatially symmetric, so for fermions it reaches
flavor-antisymmetric pairs alone.  The fully $S^z=0$ disordered
state is flavor-symmetric and has no weight there, so its energy is exactly $(4N-6)V_1$ and independent of $V_0$, whereas the ordered state does have
weight through its $S^z=\pm1$ admixture.  Raising $V_0$ therefore penalizes
the ordered phase alone and pushes $D_c$ down, against the rise in overall
scale that would push it up, and the two balance at the turnover.

Second, $\chi$ has a maximum on the critical manifold rather than a minimum,
because the sign of $\chi=E_{3/2}/E_0$ follows that of the ground-state
energy density. This is shown in Fig.~\ref{fig:o2map}, which also indicates that the reduction to $\chi$ is needed as the ground-state energy surface itself is smooth and featureless until the nonanalytic piece is extracted from it. The above denominator is not universal, and
Ref.~\onlinecite{Wiese2026} proposes normalizing $E_{3/2}$ by a gap instead,
either to the stress tensor or to the lowest parity-odd state.  On the fuzzy sphere those gaps sit in exactly
the sectors the exact-$(L,Q)$ scheme already resolves, so both alternatives
cost nothing beyond the emulators already built, and all three
normalizations return the same stationary point and critical lines agreeing
to a few parts in $10^3$ across the zoom.

The curvature of $\chi$ about that stationary point carries the corrections
to scaling, whose stiff exponent is stable across successive windows but
some $18\%$ above the XY target, while the soft one is unconverged.  The
Ising methodology validated in Sec.~\ref{sec:ising} behaves the same way at these window
sizes and settles only several sizes later, so converged O(2) exponents
require sizes beyond the present three-flavor ceiling. \label{sec:sweet}The curvature is strongly anisotropic, and its stiff direction lies transverse to the line, which is
therefore well determined and shifts by at most $0.045$ in $D_c$ when the
truncation order is changed; its soft direction lies along the line, where
the stationary point drifts with $N^{\max}$ without arresting and, under the
same change, moves by several tenths in $V_0$ or vanishes.
  Scoring the predicted dimensions against the bootstrap along the continuous
finite-size line, we find in both models that the deviation has a clear minimum, and that this minimum lies away from the $\chi$ point.

\section{Summary and outlook}
\label{sec:concl}

This work introduces a computationally efficient framework for exploring the coupling space of fuzzy-sphere models in search of their conformal critical points. We formulate an ED approach that involves symmetry-adapted bases with exact total angular momentum $L$, as opposed to the commonly used $L_z$-bases with a well-defined projection only. Besides providing clean symmetry-resolved spectra (with no a posteriori $\langle L^2\rangle$ assignments), the exact-$L$ blocks of reduced dimensionality extend the reach of ED to larger system sizes than those accessible in the $L_z$ scheme.

We then show that a subspace spanned by a few ED solutions is enough to precisely capture the states' behavior across entire regions in the space of couplings. The efficiency of such reduced-basis emulators relies on the low dimensionality of the manifold swept by the states as the couplings are varied, which is specific to finite-size systems such as the ones considered here. The parameter dependence of the various RBM observables is continuous and it therefore removes the grid interpolation step to which various subsequent analyses may be sensitive. In particular, the RBM emulators also provide analytic parameter derivatives, which enter directly the identification of the fuzzy-sphere models' critical couplings. 

Upon confirming that the exact-$L$ ED + RBM approach introduced here successfully reproduces the data of Refs.~\onlinecite{Zhu2023, Wiese2026} regarding the determination of the critical line for the Ising model by finite-size scaling and ground-state energy criteria, we extended the analysis to the O(2) model of Ref.~\onlinecite{Dey2026a}. The critical lines found by both criteria agree well with each other and with the locations of the critical points determined in that work. Both criteria rest on ground state data alone, with no input from conformal field theory at any stage, and for the O(2) model the energy criterion comes within half a percent of a DMRG determination carried out at more than twice the system size.  We further find that the point a ground-state energy criterion selects along that line is determined far less sharply than the line itself, since the curvature that fixes it is much softer along the line than across it. The couplings at which the computed dimensions best match the bootstrap do not coincide with that point. 

The exact-$L$ ED construction extends to higher-dimensional fuzzy quantum models \cite{Meng2026}, with their enlarged symmetries and parameter spaces. On the fuzzy three-sphere the rotation algebra splits into two commuting $\mathfrak{su}(2)$ factors, so the exact-$L$ construction generalizes to bases that carry both angular momenta exactly. Resolving two angular momenta rather than one compounds the reduction in block size, which already exceeds two orders of magnitude at the largest size diagonalized in Ref.~\onlinecite{Meng2026} and grows with the system's size. It also labels the $SO(4)$ content of every level without the a posteriori Casimir assignment on which the operator matching relies there. The interaction is again affine in its pseudopotentials and fields, so the emulators carry over unchanged to a five-dimensional coupling space beyond the reach of a dense scan. 

Finally, the RBM approach may be generalized to include excited states, see e.g. Appendix~\ref{app:ope}, so criteria that use the spectrum can
replace the ground-state ones used here.
The solver used to obtain the snapshots need not
necessarily be ED either, and adopting a DMRG solver \cite{Brehmer2023Aug, Baran2023Apr} would enable fast  parameter space scans for large systems beyond the reach of ED.

\begin{acknowledgments}
Early discussions with M. Burrello about the fuzzy sphere regularization are kindly acknowledged. This work was supported by a grant of the Romanian Ministry of
Research, Innovation and Digitization, project number PN-23-21-01-01/2023.
The exact-$(L,Q)$ diagonalization and certified reduced-basis codes used
here are freely available as the Julia package FuzzySphereLDiag
\cite{FuzzySphereLDiag}, together with a tutorial reproducing the
cross-code checks against FuzzifiED \cite{Zhou2025}.
The exact-diagonalization and reduced-basis codes, the numerical
validation suite, the figures, and portions of the text were developed
with the assistance of the AI assistant Claude (Fable 5 and Opus 5, Anthropic),
working under the author's direction.  All results were verified by the
author, who bears sole responsibility for the content.
\end{acknowledgments}

\appendix

\section{Validation against published O(2) conformal data}
\label{app:ope}

Conformal perturbation theory writes the shift of each level from the
vacuum as
\begin{equation}
\delta E_o(D)=u(D)\,\Delta_o+4\pi g_\varepsilon(D)\,f_{o\varepsilon o},
\label{eq:cpt}
\end{equation}
with $u=v/R$ the nonuniversal energy scale of Sec.~\ref{sec:methods} and
$g_\varepsilon$ the coupling to the relevant scalar.  With $\Delta_\sigma=0.519088$ and
$f_{\sigma\varepsilon\sigma}=0.687126$ taken from the bootstrap
\cite{Chester2020}, the pair $\sigma$, $\partial_\mu\sigma$ fixes $u$ and $g_\varepsilon$ at every $D$, where the descendant carries $f_{\partial\sigma\varepsilon\partial\sigma}=f_{\sigma\varepsilon\sigma}
[1+\Delta_\varepsilon(\Delta_\varepsilon-3)/6\Delta_\sigma]$; the critical
coupling is the root of $g_\varepsilon$, and every other
$f_{o\varepsilon o}$ follows from the slope of its own level there, a derivative the emulator returns in closed form instead of by differencing.

The extraction of Ref.~\onlinecite{Dey2026a} runs here on emulators for the
six sectors that carry the
operators involved, $(L,Q)=(0,0)$, $(1,0)$, $(2,0)$, $(0,1)$, $(1,1)$ and
$(0,2)$, at $N=8$--$12$: $395$ snapshots in all, $11$--$16$ per sector,
standing for blocks of dimension up to $1.9\times10^5$.  One reduced
eigensolve returns both $\delta E_o$ and $\partial\delta E_o/\partial D$,
the latter as $y^\dagger\mathsf{h}_2\,y$ by Hellmann--Feynman,
Eq.~\eqref{eq:hf}, with $\mathsf{h}_2$ the projected affine piece
conjugate to $D$, so Eq.~\eqref{eq:cpt} is evaluated in closed form at any $D$.

The critical couplings reproduce those of Ref.~\onlinecite{Dey2026a} to every
digit it prints, $D_c=2.8747$, $2.8516$, $2.8347$ and $2.8123$ at $N=8$,
$9$, $10$ and $12$, with $2.8220$ at the $N=11$ it does not list.  The OPE
coefficients agree to $1$--$4\times10^{-4}$ (Table~\ref{tab:ope}), the
level at which its values are quoted, and the same solves return the
dimensions $\Delta_\varepsilon=1.506$, $\Delta_{j_\mu}=1.960$,
$\Delta_{T_{\mu\nu}}=3.017$ and $\Delta_t=1.283$ at $N=12$, against the
exact $2$ and $3$ for the two conserved currents.

The agreement spans a charge-neutral scalar, a conserved current, the
stress tensor and a charge-two scalar.  Both calculations read the same
microscopic model with the same estimator, but obtain the parameter
derivatives the estimator rests on in entirely different ways.  The
residual gap to the bootstrap, largest for $\varepsilon$ at $3\%$, is a
finite-size effect shared by both and discussed in
Ref.~\onlinecite{Dey2026a}.

\begin{table}[ht!]
\caption{\label{tab:ope}Diagonal OPE coefficients $f_{o\varepsilon o}$ of the
O(2) Wilson--Fisher CFT at $V_0=4$, extracted from the emulators of
Sec.~\ref{sec:rbm} by the procedure of Ref.~\onlinecite{Dey2026a} and compared
with the exact-diagonalization column of its Table~2, taken at the same
system size $N=12$, and with the conformal bootstrap
\cite{Chester2020}.  The $\sigma$ row is not a test: $f_{\sigma
\varepsilon\sigma}$ is the bootstrap input that calibrates the
extraction.}
\begin{ruledtabular}
\begin{tabular}{lcccc}
$o$ & $(S_z,L)$ & this work & Ref.~\onlinecite{Dey2026a} & bootstrap \\
\hline
$\varepsilon$    & $(0^+,0)$ & $0.8565$ & $0.856146$ & $0.830914(32)$\\
$j_\mu$          & $(0^-,1)$ & $0.9760$ & $0.976163$ & $0.9674(60)$\\
$T_{\mu\nu}$     & $(0^+,2)$ & $0.5734$ & $0.573559$ & ---\\
$t$              & $(2,0)$   & $1.2560$ & $1.256118$ & $1.25213(14)$\\
$\sigma$         & $(1,0)$   & \multicolumn{3}{c}{$0.687126$, bootstrap input}\\
\end{tabular}
\end{ruledtabular}
\end{table}

\bibliography{refs}

\end{document}